\documentclass{aa} 

\usepackage{graphicx}
\usepackage{txfonts}

\begin{document}

   \title{A new type-1 transition in NGC~1346 revealed by (spectro)polarimetry}

   \author{F. Marin\inst{1}          
          \and
          D. Hutsem\'ekers\inst{2}   
          \and
          S. Jorstad\inst{3}  
          }

   \institute{Universit\'e de Strasbourg, CNRS, Observatoire Astronomique de Strasbourg, UMR 7550, 11 rue de l'universit\'e, 67000 Strasbourg, France\\
             \email{frederic.marin@astro.unistra.fr}
             \and            
                 Institut d’Astrophysique et de G\'eophysique, Universit\'e de Li\`ege, All\'ee du 6 Ao\^ut 19c, B5c, 4000 Li\`ege, Belgium
             \and            
                 Institute for Astrophysical Research, Boston University, 725 Commonwealth Avenue, Boston, MA 02215, USA
             }

   \date{Received April 16, 2025; accepted July 6, 2025}

   \abstract 
{We present the first photo- and spectropolarimetric observations of the changing-look active galactic nucleus NGC~1346 taken with the Perkins telescope (November 2022, March 2025) and VLT/FORS2 (August and September 2024), which have revealed a new spectral transition. We find that NGC~1346 has reverted to its former (2001) type-1 spectral state after having been a type-2 active galactic nucleus from 2004 to 2022. More and more intense broad H$\beta$ (3720 $\pm$ 310 km.s$^{-1}$ full width at half maximum) and H$\alpha$ (2985 $\pm$ 230 km.s$^{-1}$) emission lines appeared between August and September 2024, suggesting that the broad-line region (BLR) is increasingly irradiated. The strong blueshift of the Balmer lines ($\sim$ 1600 km.s$^{-1}$) indicates that the gas is moving toward the observer, possibly due to a radiation-driven outflow. The strong difference between 2022's ($\sim$ 3\%) and 2024-2025's de-biased polarization ($\le$ 0.5\%), the blueing of the spectra between August and September 2024, and the slow rise in the integrated flux between 2022 and 2025 argue against asymmetric or temporary obscuration by a cloud passing in front of the line of sight. Two options, either a re-illumination of the BLR or a tidal disruption event, can explain the observed properties of NGC~1346. Timely follow-up observations taken while the phenomenon is still ongoing are needed to determine which of the two solutions is the correct one.}


   \maketitle
%

\section{Introduction}
\label{Introduction}
Active galactic nuclei (AGNs) are powered by accretion onto a supermassive black hole (SMBH); this produces multiwavelength emission that varies with the accretion rate and surrounding structure \citep{Sartori2018}. At the heart of AGN unification models lies a key observational distinction: type-1 AGNs show broad Balmer emission lines from the fast-moving, equatorial broad-line region (BLR), while type-2 AGNs do not, as the BLR is hidden from direct observation by a dusty circumnuclear structure — the so-called torus.— What drives this distinction is, at first order, the inclination of the AGN: objects seen pole-on reveal their BLR in direct light through the torus funnel, while the BLR emission lines of   AGNs seen edge-on are blocked by the equatorial torus \citep{Antonucci1993}. Because AGNs are parsec-scale objects, their inclination with respect to an observer (and thus their spectral classification) should not change on human timescales, yet some AGNs exhibit dramatic spectral transitions between type-1 and type-2 states.

NGC~7603 was the first reported case of an AGN type variation occurring in less than a year \citep{Tohline1976}. Between November 6, 1974, and November 8, 1975, its broad Balmer lines disappeared. This would mean that the AGN switched from type-1 to type-2 in a year, which is physically challenging to explain. The same kind of appearance and/or disappearance was observed for NGC 4151 \citep{Penston1984}, Mrk 1018 \citep{Cohen1986}, Mrk 590 \citep{Denney2014}, and NGC 1346 \citep{Senarath2019}. Although "changing-look" AGNs  were discovered almost 50 years ago, the question of what triggers  them remains. 

Several scenarios have been proposed. Type variation could be due to (1) opaque clouds crossing the line of sight \citep{Risaliti2007}, (2) dramatic changes in the ionizing continuum that can generate changes in the structure of the BLR or torus \citep{Noda2018}, (3) supernova outbursts \citep{Aretxaga1999}, or (4) tidal disruption events \citep{Merloni2015}. The first scenario works well to explain X-ray variations in fluxes, resulting in a transition between a transmission- and a reflection-dominated spectrum, but this requires very large clouds to obscure the entirety of the accretion disk optical emission \citep{Risaliti2007}. The second scenario is appealing because a change in the ionizing continuum power would result in a variation of the dust sublimation radius and, thus, a shrinking (or even a disappearance) of the BLR. Most changing-look AGNs are found near the critical luminosity below which the BLR disappears \citep{Noda2018}. A supernova can explain the appearance and sudden disappearance of a broad line but fails to explain the observed cases where the broad line reappears several years later (e.g., Mrk 590; see \citealt{Mandal2021}). Finally, changing-look AGNs could also be the results of a luminous flare produced by the tidal disruption of a super-solar-mass star passing just a few gravitational radii outside the event horizon of a SMBH. However, in general, flares in changing-look AGNs last longer than tidal disruption events \citep{Merloni2015}.

Active galactic nucleus spectral type evolution, especially if repeated in time, challenges our understanding of AGN structure and variability. Spectropolarimetry is a powerful tool that can be used to decipher this mystery as it is uniquely sensitive to changes in the emitting, scattering, and absorbing regions of AGNs. Anything that makes the geometry of the target change, whether BLR shrinking, clouds in the light of sight, or additional sources of emission, will show up in polarized light \citep{Marin2020}. For this reason, we undertook observation of a sample of changing-look AGNs with the FOcal Reducer/low dispersion Spectrograph 2 (FORS2) mounted on the Very Large Telescope (VLT). Of the sample, to be presented in another publication, NGC~1346 had the most unexpected behavior, which motivated the present paper.

NGC~1346 is a spiral galaxy situated in the Eridanus constellation (right ascension 52.555326$^\circ$, declination -5.543323$^\circ$), at a redshift of $z$ = 0.013539 \citep{Paturel2003}. In the standard $\Lambda$ cold dark matter cosmology, this corresponds to a Hubble distance of $\sim$ 57.61~Mpc. NGC~1346, whose SMBH mass was estimated to be on the order of 2.4 $\times$ 10$^7$ solar masses \citep{Gureev2010}, was first cataloged as a type-1 AGN by \citet{Veron2003}, but its optical spectral type changed with time, evolving to a type-2 classification \citep{Senarath2019} due to the disappearance of its broad Balmer lines. This qualifies NGC~1346 as a changing-look AGN. Section~\ref{Observation} presents our new observations, and Sect.~\ref{Analysis} analyzes our findings in the context of the unified model of AGNs. We discuss and conclude our work in Sect.~\ref{Conclusions}.

\section{Observations}
\label{Observation}

\subsection{VLT/FORS2}
\label{Observation:FORS2}

Spectropolarimetric observations of NGC~1346 were obtained using the European Southern Observatory (ESO) VLT/FORS2 mounted at the Cassegrain focus of Unit Telescope \#1 (Antu). Linear spectropolarimetry was performed by inserting a Wollaston prism in the beam that splits the incoming light rays into two orthogonally polarized beams separated by 22" on the charge-coupled device (CCD) detector. To derive the normalized Stokes parameters $u$ and $q$, two consecutive series of frames were obtained at a given epoch with the half-wave plate rotated at four different position angles: 0$^\circ$, 22.5$^\circ$, 45$^\circ$, and 67.5$^\circ$. This combination allowed us to remove most of the instrumental polarization. The polarization data were extracted independently for the two series of frames. They do not show any significant difference, which means that the ordinary-to-extraordinary beam flux ratio did not change significantly between the two cycles. To further check this, we computed the depolarization factor following \citet{Afanasiev2012} and \citet{Jiang2021}, for each half-wave plates (HWP) position. This factor does not vary by more than 0.1\% between the two cycles. We then only considered, for each epoch, the average of the derived Stokes parameters.

The spectra were acquired with the grism 300V and the order-sorting filter GG435 (4600-8650 \AA). Observations were carried out on August 12 and September 4, 2024, with 3200 s of total integration time each. The slit width was 1" on the sky, providing us with an average resolving power of R $\approx$ 440 around 5849~\AA, enough to correctly sample the broad emission lines. The slit was positioned along the north-south direction. CCD pixels were binned 2 $\times$ 2, which corresponds to a spatial scale of 0.25" per binned pixel. The airmass was between 1.4 and 1.1, the sky clear, and the seeing around 1.0" in August and 0.7" in September. A polarized standard star (Hiltner 652, \citealt{Fossati2007}) was observed on July 26 and August 28, and an unpolarized standard star (WD 1620-391) on August 28. The instrumental polarization estimated from the unpolarized standard stars is lower than 0.1\%.

Raw frames were first processed to remove cosmic-ray hits using the Python implementation of the "lacosmic" package \citep{Dokkum2001,Dokkum2012}. The ESO FORS2 pipeline \citep{Izzo2019} was then used to obtain images with two-dimensional spectra flat-fielded, rectified, and calibrated in wavelength. The one-dimensional spectra were extracted using an extraction aperture of 2.75" (11 pixels), which is roughly 2.5 times the highest seeing value. The aperture for the extraction of the FORS2 spectra was chosen as small as possible to minimize the host contamination, given the seeing. The sky spectrum was estimated from adjacent multi-object spectroscopy (MOS) strips and subtracted from the nucleus spectrum. The normalized Stokes parameters $u$ and $q$ were then computed from the ordinary and extraordinary spectra according to the procedure described in the FORS2 manual\footnote{https://www.eso.org/sci.html}, and corrected for the half-wave plate chromatic dependence. The total flux was corrected for the atmospheric extinction and calibrated using a master response curve. Uncertainties were estimated by propagating the photon and readout noises. The polarization degree, $P,$ and polarization angle, $\theta,$ were finally computed using the standard formulae:

\begin{equation}
P = \sqrt{q^{2}+u^{2}},
\end{equation}
and 
\begin{equation}
\theta = \frac{1}{2}\arctan\,(\frac{u}{q}) +\Delta ,
\end{equation}
where $\Delta = 0^\circ$ for $u > 0$ and $q > 0$; $90^\circ$ for $q < 0$; and $180^\circ$ for $u < 0$ and $q > 0$, according to the usual north-south convention (0$^\circ$ in the north, 90$^\circ$ in the east). All reported polarization values, including the ones obtained with the Perkins telescope (see Sect.~\ref{Observation:Perkins}), are de-biased following \citet{Simmons1985} recommendations: $P_d = \sqrt{P^2-{\sigma_P}^2}$, with $\sigma_P$ the uncertainty on the polarization degree, $P$.

To estimate the importance of the host starlight contamination, we changed the extraction aperture of the FORS2 spectra to 5" and 10". The polarization measured with these apertures is nearly unchanged ($\Delta P <$ 0.05\% and $\Delta \theta <$ 8$^\circ$), indicating that dilution by the host galaxy light already affects the spectra extracted with the smallest aperture, as well as a significant contribution of the interstellar polarization (ISP) in our Galaxy. The ISP contamination was measured using the star GaiaDR2 3244105248118361984 (RA = 54.193032$^\circ$, dec = -7.121268$^\circ$), situated about 2.27 degrees from the AGN. The polarization recorded toward this star is $P$ = 0.25\% $\pm$ 0.10\% at 39.1$^\circ$ $\pm$ 12.8$^\circ$ \citep{Panopoulou2025}. It was subtracted from our VLT/FORS2 data.

\subsection{Perkins telescope}
\label{Observation:Perkins}

\begin{table*}
\caption{Perkins telescope data.}              
\centering                                     
\begin{tabular}{l c c c c c c c }          
\hline\hline                        
Date & Band & Magnitude & q (\%) & u (\%) & $P_d$ (\%) & $\theta$ ($^\circ$) & Air mass\\   
\hline                                   
Nov. 23, 2022 & B  & 15.965  $\pm$ 0.038  & -0.91  $\pm$ 1.15  & -2.83  $\pm$ 1.15  & 2.73  $\pm$ 1.15  & 126.1  $\pm$ 11.1 & 1.46-1.59 \\
Nov. 23, 2022  & V  & 15.119  $\pm$ 0.041  & 1.18  $\pm$ 0.92  & -2.27  $\pm$ 0.91  & 2.38  $\pm$ 0.91  & 148.7  $\pm$ 10.3 & 1.38-1.45 \\
Nov. 23, 2022  & R  & 14.504  $\pm$ 0.024  & 2.02  $\pm$ 0.86  & -2.31  $\pm$ 0.86  & 2.95  $\pm$ 0.86  & 155.6  $\pm$ 8.0 & 1.34-1.37 \\
Nov. 23, 2022  & I  & 13.904  $\pm$ 0.033  & 1.69  $\pm$ 0.82  & -1.60  $\pm$ 0.82  & 2.14  $\pm$ 0.82  & 158.3  $\pm$ 10.1 & 1.60-1.68 \\
Mar. 25, 2025  & R  & 14.242  $\pm$ 0.009  & 0.50  $\pm$ 0.35  & -0.22  $\pm$ 0.35  & 0.42  $\pm$ 0.35  & 168.1  $\pm$ 18.2 & 2.60-2.97 \\
\hline                                            
\end{tabular}
\label{Tab:Perkins}      
\end{table*}

Prior to and after our VLT/FORS2 observation, we obtained polarimetric and photometric observations of NGC 1346 at the 1.83~m Perkins telescope (Perkins Telescope Observatory, Flagstaff, AZ, USA) using the PRISM camera equipped with a polaroid (POL-HN38) and a wheel of UBVRI filters with $\lambda_{eff}$ equal to 3663~\AA, 4361~\AA, 5448~\AA, 6407~\AA, and 7980~\AA, respectively. A mounted piece of HN38 Polaroid is located near an internal pupil plane in a filter wheel cell. The polaroid cell is rotated to fixed angles between exposures under instrument control. The polarimetric observations were carried out in the B,V, R, and I bands on November 23, 2022, while on March 25, 2025, another polarimetric observation was performed in the R band. 

Each polarimetric observation involves three series of Stokes parameters (I, q, and u measurements) and a series of four observations of the polaroid, at instrumental position angles of 0$^\circ$, 90$^\circ$, 45$^\circ$, and 135$^\circ$ and with an exposure of 180, 150, 150, and 120~s for the B, V, R, and I bands, respectively, at each position angle. The q and u parameters were averaged over the series to calculate the degree of polarization, $P$, and position angle of polarization, $\theta$, and their uncertainties. We used a circular aperture with a diameter of 10'' to measure the core of NGC 1346, stars in the field, and the sky background. Despite the difference in apertures between the Perkins telescope and the VLT/FORS2 data, the measured polarization remains comparable since the host contamination is basically aperture-independent, as seen in Sect.~\ref{Observation:FORS2}. Finally, the images were corrected for bias and flat field. During the observations in November 2022 and March 2025, the seeing was $\sim$2", which is typical for the Perkins telescope site.

Since the camera has a wide field of view (14' $\times$ 14'), we used field stars to perform both interstellar and instrumental polarization corrections, assuming that stars in the field are intrinsically unpolarized. We used unpolarized calibration stars from \cite{Schmidt1992} to check the instrumental polarization, which is usually within 0.2\%, and polarized stars from the same paper to calibrate the polarization position angle. If we ignore instrumental polarization due to its small values, a level of the ISP in the vicinity of NGC1346 is $\sim$0.5\%. We employed a differential photometry method to obtain photometry of the target using stars in the field of NGC 1346. Results from our observational campaign are reported in Table~\ref{Tab:Perkins}.

\section{Analysis}
\label{Analysis}

\begin{figure*}
\centering
\includegraphics[width=\textwidth]{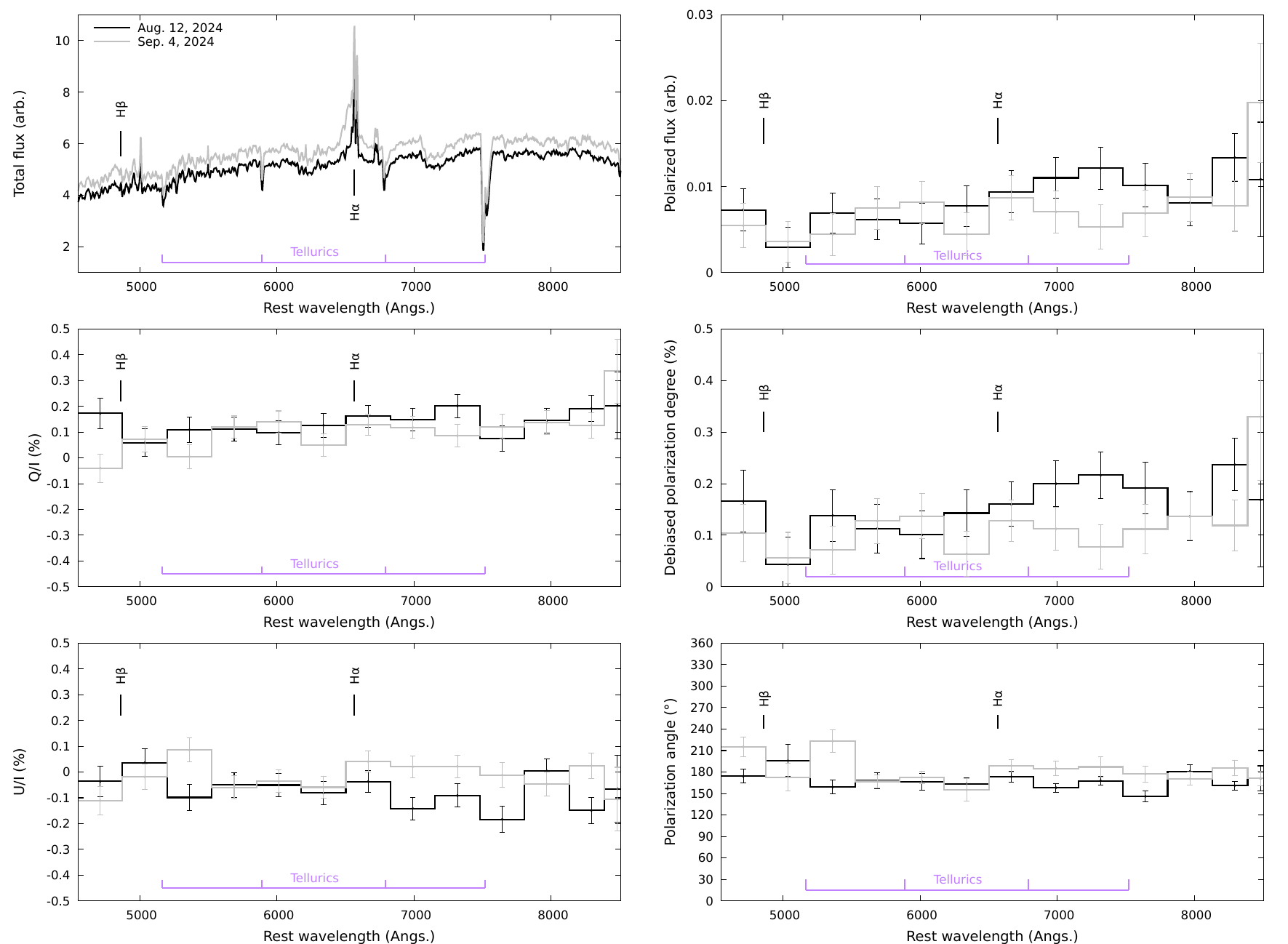}
\caption{NGC~1346 observation taken with the VLT/FORS2 on August 12, 2024 (solid black line) and on September 4, 2024 (solid gray line). Top-left: Total flux spectra (in arbitrary units, but not renormalized). Middle- and bottom-left: $q$ (= Q/I) and $u$ (= U/I) normalized Stokes parameters. Top-right: Polarized fluxes, that is, the total flux multiplied by the polarization degree. Middle-right: De-biased linear polarization degrees. Bottom-right: Polarization position angles. Except for the total flux panel, spectra were re-binned to 100 consecutive pixels. Observational errors are indicated for each spectral bin, and the locations of H$\alpha$ and H$\beta$ emission lines, as well as the strongest atmospheric absorption bands (tellurics), are clearly indicated.}
\label{Fig:Binned_data}%
\end{figure*}

\begin{figure}
\centering
\includegraphics[width=\columnwidth]{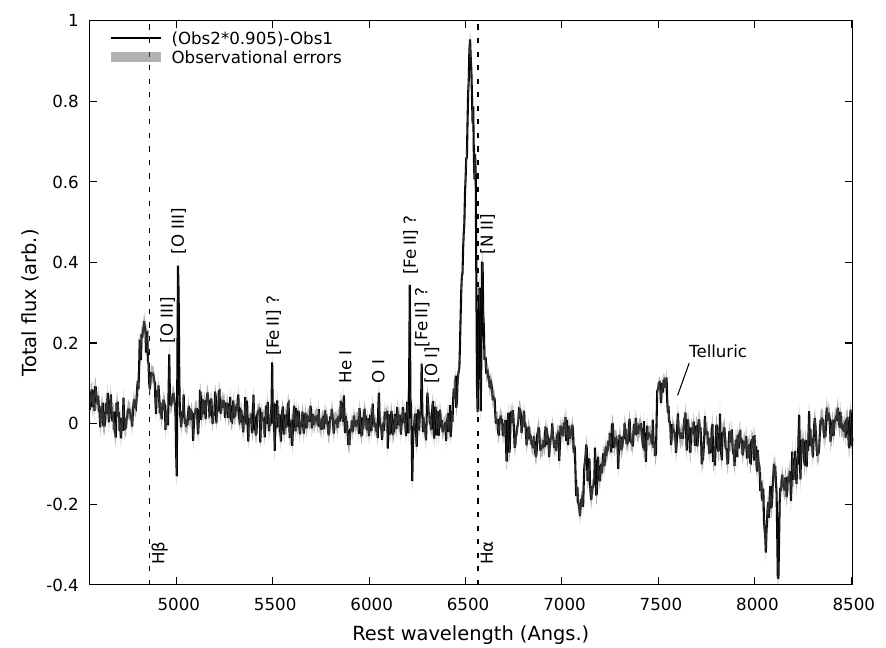}
\caption{Flux difference between the two observations of NGC~1346: Obs1 (August 12) and Obs2 (September 4). The Obs2 continuum was normalized to the Obs1 continuum by multiplying Obs2's total flux by 0.905, before the two spectra were subtracted. The centroid of the H$\alpha$ (6563~\AA) and H$\beta$ (4861~\AA) emission lines in air are indicated. A feature due to telluric absorption, which could be misinterpreted as an emitted broad line, is indicated. Observational errors are indicated in transparent gray.}
\label{Fig:Difference}%
\end{figure}

\subsection{Spectroscopy}
\label{Analysis:Spectra}

The spectra of the two epochs are illustrated in Fig.~\ref{Fig:Binned_data}, binned in wavelength for the polarization signal. Focusing on the total flux spectra first (top-left panel of Fig.~\ref{Fig:Binned_data}), we can see that the two spectra are very similar, with moderately reddened continua and bright emission lines. At first glance, the spectra appear similar. However, a notable difference emerged: the blue wing of the H$\alpha$ line appears much larger in September with respect to the August spectrum. To verify this, we normalized the two spectra by matching their continuum fluxes before subtracting them, which results in the difference spectrum shown in Fig.~\ref{Fig:Difference}. 

By doing so, the evolution of NGC~1346 in less than a month becomes blatant. Not only the H$\alpha$ emission line exhibit a newly emerged, broad and blueshifted component, but H$\beta$ too. Most narrow emission lines appear nearly identical between the two epochs and therefore almost cancel out in the difference spectrum. Small residuals can still be seen at the positions of several narrow lines (indicated in Fig.~\ref{Fig:Difference}), but they are significantly sharper than their counterparts in the original spectra, often spanning just a few spectral bins. This suggests that they are likely not physical but instead arise from slight mismatches in wavelength calibration, observing conditions, spectral resolution, or residual noise in the subtraction process. The difference spectrum also shows minor telluric residuals unrelated to the AGN itself.

\begin{figure*}
\centering
\includegraphics[width=\columnwidth]{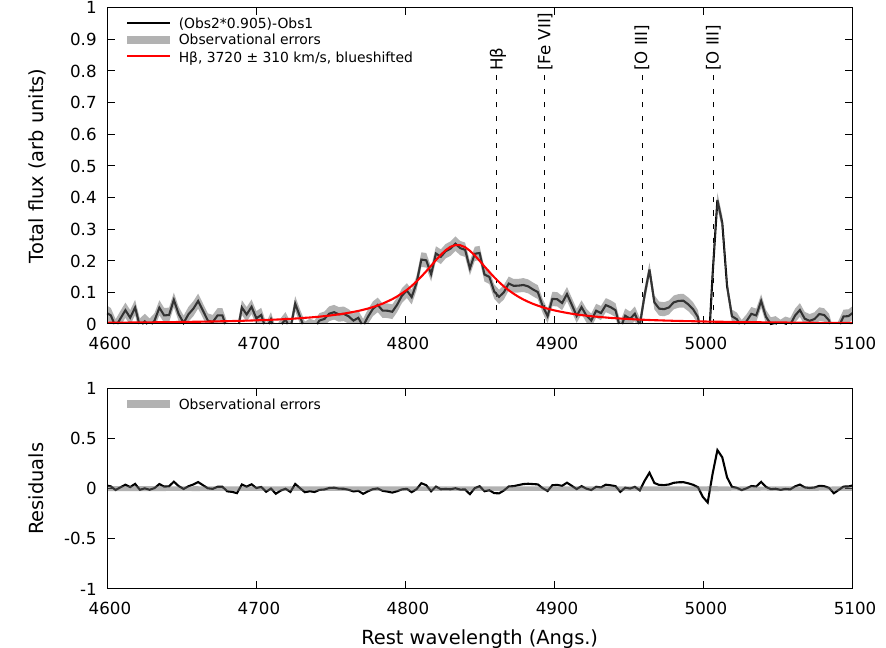}
\includegraphics[width=\columnwidth]{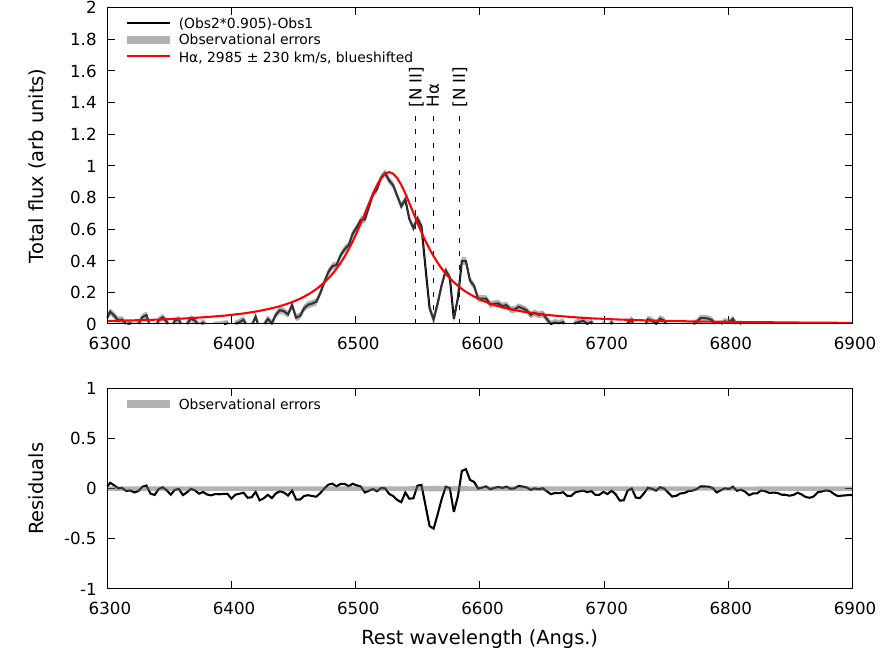}
\caption{Top: Lorentzian fits of the newly appeared broad emission features in the September spectrum. The fitted data are the ones presented in Fig.~\ref{Fig:Difference}. The centroid wavelength (in air) of various emission lines is indicated. Bottom: Residuals. Observational errors are indicated in transparent gray. The left column is for the H$\beta$ emission line, while the right column shows the fits of the H$\alpha$ feature.}
\label{Fig:Fits}%
\end{figure*}

Fitting the newly appeared broad and blueshifted lines with Lorentzian profiles (similar results are obtained with Gaussian fits) yields full width at half maximum (FWHM) values of 2985 $\pm$ 230 km.s$^{-1}$ for H$\alpha$ and 3720 $\pm$ 310 km.s$^{-1}$ for H$\beta$; see Fig.~\ref{Fig:Fits}. Due to these broad components, we  classify NGC~1346 as a type-1 AGN, in contrary to what was observed in 2018 by \citet{Senarath2019,Senarath2021}. We also note that both H$\alpha$ and H$\beta$ lines are strongly blueshifted with respect to their usual emission wavelength in air. The H$\alpha$ velocity shift is -1635 $\pm$ 228 km.s$^{-1}$, while it is -1622 $\pm$ 308 km.s$^{-1}$ for the H$\beta$ feature. Finally, we note that the continuum level is not perfectly equal to zero and shows a residual blue component. 

Focusing on the narrow emission lines, we find that they are consistent with emission in a lower-density region with inhomogeneities. The narrow lines present FWHM inferior to 1000 km.s$^{-1}$, as expected from typical AGNs. From the line intensities provided by our fits, computing the emission-line intensity ratios log$_{\rm 10}$([N II]$_{\rm 6583~\AA}$ / H$\alpha_{\rm 6563~\AA}$) versus log$_{\rm 10}$([O III]$_{\rm 5007~\AA}$ / H$\beta_{\rm 4861~\AA}$) places NGC~1346 deep in the AGN region of the Baldwin--Phillips--Terlevich (BPT) diagnostic diagram \citep{Baldwin1981,Veilleux1987}, specifically in the Seyfert quadrant (see Sect.~\ref{Conclusions} for additional details about this classification over various epochs.)

\subsection{Polarization}
\label{Analysis:Polarization}

Spectropolarimetric campaigns of changing-look AGNs are scarce, and very few have been achieved shortly after the spectral type transition of an object (see \citealt{Hutsemekers2020} for an example). We show in Fig.~\ref{Fig:Binned_data} the polarization of NGC~1346 at the two epochs. Because of the insufficient signal-to-noise ratio in polarization (about 0.36 per spectral element), we had to bin the data every 100 pixels in wavelengths to reach a signal-to-noise ratio of 3.6, which corresponds to 330~\AA\, per bin. 

The observed binned polarization is really small, below half a percent most of the time. This is likely due to the host galaxy that emits unpolarized starlight, diluting the intrinsic polarization of the AGN, on top of which ISP leaves its imprint. There is, unfortunately, no known template for the host spectrum of NGC~1346 to correct for this diluting effect. The ISP-corrected polarization we observe is, apparently, wavelength-independent (within the errors bars). This would suggests that electron scattering is the dominant process by which polarization arises. The polarization angle is too noisy to determine if it is also wavelength-independent. Integrated over the whole observed waveband, $P_d$ = 0.15\% $\pm$ 0.01\% and $\theta$ = 165.7$^\circ$ $\pm$ 2.6$^\circ$ in August, and $P_d$ = 0.09\% $\pm$ 0.01\% and $\theta$ = 175.8$^\circ$ $\pm$ 4.1$^\circ$ in September (uncertainties on the polarization degree and the polarization angle were derived following standard error propagation and computed from the uncertainties in the normalized Stokes parameters $q$ and $u$). It suggests that the overall polarization decreased within a month, associated with a small rotation of the polarization position angle. Alas, our signal-to-noise ratio is insufficient to detect subtle spectral variations in $P_d$ that could tell us if ISP has been sufficiently corrected and confirm the presence or absence of polarized broad lines with confidence.

Fortunately, the source was observed prior to this spectral transition -- in November 2022 -- by the Perkins telescope (see Table~\ref{Tab:Perkins}). The B, V, R, and I photo-polarimetric observation show that $P_d$ was much higher in 2022 (about 3\%) than at the time the VLT observed NGC~1346. Such a high linear polarization degree is typically what is expected from type-2 AGNs in the optical and near-infrared bands, where starlight dilution strongly impacts the observed polarization signatures \citep{Marin2014}. R-band polarimetric data taken with the same telescope in March 2025 -- after the VLT/FORS2 campaign -- reveals that the source polarization is well below its 2022 values, without statistically significant variations of $\theta$. In other words, the Perkins polarization data reveal that NGC~1346 had a level of polarization consistent with that of typical type-2 AGNs in November 2022 \citep{Marin2014}, while the VLT/FORS2 data demonstrate that the source became a type-1 AGN somewhere between November 2022 and August 2024, in terms of both spectroscopy and polarimetry. The recent (March 2025) polarimetric acquisition by the Perkins telescope tends to confirm that the source is still in a type-1 state, polarimetrically speaking.

While AGNs usually exhibit high optical linear continuum polarization ($\gg$1\%) perpendicular to the radio jet axis in type- 2 objects, and low polarization ($\le$1-2\%), often aligned with the jet axis, in type-1 objects, the polarization measured in changing-look AGNs, in either a type-1 or type-2 state, was found to be very small \citep{Hutsemekers2017,Hutsemekers2019,Hutsemekers2020}. This led to the conclusion that the disappearance of the broad emission lines is not due to dust obscuration, but rather to a change in the accretion rate, as confirmed by numerical simulations \citep{Marin2019,Marin2020}. The high polarization detected in NGC~1346 in its type-2 state thus indicates that the change of state is due either to variable obscuration or to the echo of a previous type-1 phase \citep{Hutsemekers2019,Marin2020}, the latter scenario being more likely (see the next section). It is regrettable that the position angle of the jet structure in the AGN is unknown\footnote{\cite{WU2023} present multifrequency radio images of NGC~1346. Although the contours of the FIRST (Faint Images of the Radio Sky at Twenty–cm) image appear elongated along a position angle of $\sim$ 145$^\circ$, this elongation is not observed in the higher resolution VLASS (Very Large Array Sky Survey) image, casting doubt on its reality.}, so we cannot determine if NGC~1346 shows the usual behavior of type-1 objects (polarization angle parallel to the radio jet axis) or the behavior of type-2s (polarization angle perpendicular to the radio jet axis).

\section{Discussion and conclusions}
\label{Conclusions}

We have discovered that NGC~1346 changed back to its former (2001) type-1 spectral classification since its last spectroscopic observation in 2018 \citep{Senarath2019}. Our Perkins polarimetric measurements even support the hypothesis that the source was still in a type-2 spectral state in November 2022. VLT/FORS2 spectra taken in 2024 show that the AGN now has a type-1 classification, with more and more intense broad (3720 $\pm$ 310 km.s$^{-1}$ and 2985 $\pm$ 230 km.s$^{-1}$), blueshifted (-1622 $\pm$ 308 km.s$^{-1}$ and -1635 $\pm$ 228 km.s$^{-1}$) H$\beta$ and H$\alpha$ emission lines (respectively) appearing between August and September. This suggests that the BLR has been getting increasingly detectable since 2022, meaning that the circumnuclear obscuration has changed in geometry and/or density, the BLR itself has become more luminous and/or extended, or a powerful flare is currently ongoing in the core of the AGN.

The fact that the Balmer lines are blueshifted suggests that a significant fraction of the BLR gas is moving toward the observer rather than moving in a purely Keplerian motion around the SMBH, as it is often considered in the literature \citep{Peterson1999}. This may be due to an outflow from the inner core of the AGN, possibly driven by radiation pressure from the accretion disk \citep{Proga2000}. An asymmetry in the BLR structure, where we would preferentially observe gas moving toward us while the receding part is either obscured or weaker, seems less plausible due to the very low polarization degree we detect, as asymmetries reinforce the observed polarization. 

\begin{figure}
\centering
\includegraphics[width=\columnwidth]{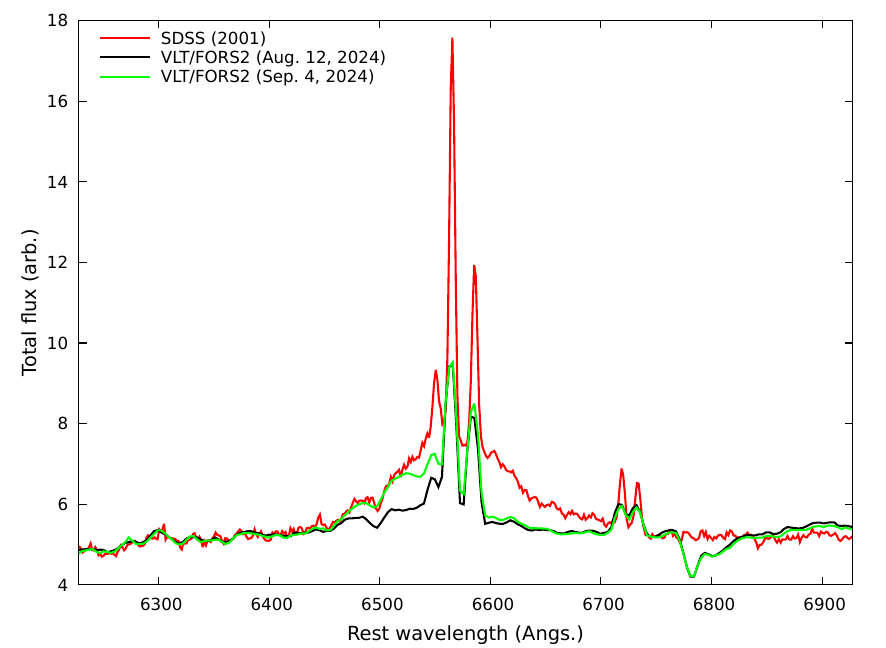}
\caption{2001 SDSS spectrum discussed in \citet{Senarath2019}, superimposed on our two VLT/FORS2 spectra (rescaled to the matching continuum). We caution the reader that the spectra are not integrated with the same aperture, so the equivalent widths of the narrow lines do not match. We show a zoomed-in view of the spectra around the broad H$\alpha$ line.}
\label{Fig:SDSS}%
\end{figure}

To check this, we retrieved the Sloan Digital Sky Survey (SDSS) spectrum from 2001 (zoomed in around the H$\alpha$ line; see Fig.~\ref{Fig:SDSS}). The computation of the BPT line ratios ([O~III]/H$\beta$ and [N~II]/H$\alpha$) shows that NGC~1346 was already located deep in the AGN region of the diagnostic diagram in 2001. Comparing the SDSS spectrum with ours is also informative: the broad SDSS line is symmetrical, suggesting that the reactivation of the BLR started with the blue wing and supporting the hypothesis of an outflow. We should therefore expect the red wing to follow by a few months, given the longer lead time. Such a behavior was already demonstrated in the case of another changing-look AGN, Mrk~1018, in which the red wing of the H$\alpha$ emission line disappeared before the blue wing as the source was evolving toward a type-2 state (i.e., the exact opposite of what we see in NGC~1346; see \citealt{Hutsemekers2020,Lu2025}). 

It is very likely that the outflow has always been present in NGC~1346, even in its type-2 period (2004 - 2022), but simply not visible due to the lack of excitation \citep{Noda2023}. Indeed, if we compute the Eddington ratio of NGC~1346 using the 2006 infrared luminosity reported by \citet{Berlind2010}, we find a value of approximately 0.024, which is below the median Eddington ratio of 0.1 for AGNs \citep{Panda2018} but still high enough to accelerate gas and produce an outflow \citep{King2015,Vivek2025}. This would explain why we observe a higher polarization and a different B-band polarization position angle in its type-2 state, where the flux scattered from the polar components  -- an echo of the type-1 state --  is stronger than the flux scattered from the under-luminous BLR \citep{Marin2020}.

Following the observed $\sim$ 0.262 magnitude decrease (i.e., flux increase) in the continuum observed between our 2022 and 2025 Perkins observations of NGC~1346, the BLR was likely re-excited by the rising continuum emission. This is why broad emission lines are slowly reappearing: the line flux intensification propagates first in the blue wing of the line with a delay of a few months (on the line of sight toward us), explaining the variations observed between August and September 2024. If this scenario is correct, we therefore expect a future flux increase in the central part of the line (in the region perpendicular to the line of sight, with relatively low speeds), then in the red wing of the line (from the region of the outflow moving away from us). Since we do not observe a 90$^\circ$ rotation of the polarization angle between 2022 and 2025, the scattering region outside the BLR has not yet been reached by the variation in the continuum flux.

If NGC~1346 is currently experiencing a continuous increase in its ionizing flux, propagating into the outflowing BLR, the H$\beta$ and H$\alpha$ line profiles should become complete and remain stable over time, accompanied by a rotation of the polarization angle. However, if the broad lines disappear quickly in the coming months, it could be the signature of a tidal disruption event. In this case, the flux in the Balmer line profile should shift from blue to the core and then to red before disappearing. The case discussed by \citet{Trakhtenbrot2019} bears a strong resemblance to that of NGC~1346. In both sources, the broad H$\alpha$ emission line becomes more intense with a delay of a few months, initially showing a high-velocity blue wing. Over time, the line profile evolves and becomes redshifted, as reported in follow-up observations by \citet{Li2022}. This spectral evolution could foreshadow the future behavior of NGC~1346 if the tidal disruption scenario is correct. In this case, the polarization of the source should change gradually before returning to its previous (type-2) value, in opposition to the first scenario, where the change in ionizing continuum and the re-excitation of the BLR would impose a 90$^\circ$ rotation of the polarization angle. Regardless, NGC~1346 is experiencing a critical phase of activity that requires rapid follow-up observations.

\begin{acknowledgements}
We would like to acknowledge the anonymous referee for her/his comments and suggestions about our paper. The research at Boston University was supported in part by National Science Foundation grant AST-2108622 and NASA Fermi Guest Investigator grant 80NSSC23K1507. This study was based in part on observations conducted using the 1.83~m Perkins Telescope Observatory (PTO) in Arizona, which is owned and operated by Boston University. DH is research director at the F.R.S-FNRS, Belgium.
\end{acknowledgements}

\bibliographystyle{aa} 
\bibliography{Bibliography} 

\end{document}